\documentclass[journal=ancac3 ]{achemso}
\usepackage{graphicx}
\usepackage{color}
\usepackage{amsmath}
\usepackage{amsfonts}
\usepackage{colortbl}
\usepackage{footnote}
\usepackage{hyperref}
\usepackage{atbegshi}
\usepackage{mciteplus}

\title{Grating-graphene metamaterial as a platform for terahertz nonlinear photonics}

\author{Jan-Christoph Deinert} \affiliation{Helmholtz-Zentrum Dresden-Rossendorf, Dresden, Germany}
\author{David Alcaraz Iranzo} \affiliation{ICFO - Institut de Ci\`{e}ncies Fot\`{o}niques, The Barcelona Institute of Science and Technology, Castelldefels (Barcelona) 08860, Spain}
\author{Ra\'{u}l P\'{e}rez} \affiliation{Catalan Institute of Nanoscience and Nanotechnology (ICN2), BIST \& CSIC, Campus UAB, Bellaterra 08193, Barcelona, Spain} 
\author{Xiaoyu Jia} \affiliation{Max-Planck-Institut f{\"u}r Polymerforschung, Ackermannweg 10, 55128 Mainz, Germany}
\author{Hassan A. Hafez} \affiliation{Fakult{\"a}t f{\"u}r Physik, Universit{\"a}t Bielefeld, Universit{\"a}tsstr. 25, 33615 Bielefeld, Germany}
\author{Igor Ilyakov} \affiliation{Helmholtz-Zentrum Dresden-Rossendorf, Dresden, Germany}
\author{Nilesh Awari} \affiliation{Helmholtz-Zentrum Dresden-Rossendorf, Dresden, Germany}
\author{Min Chen} \affiliation{Helmholtz-Zentrum Dresden-Rossendorf, Dresden, Germany}
\author{Mohammed Bawatna} \affiliation{Helmholtz-Zentrum Dresden-Rossendorf, Dresden, Germany}
\author{Alexey N. Ponomaryov} \affiliation{Helmholtz-Zentrum Dresden-Rossendorf, Dresden, Germany}
\author{Semyon Germanskiy} \affiliation{Helmholtz-Zentrum Dresden-Rossendorf, Dresden, Germany}
\author{Mischa Bonn} \affiliation{Max-Planck-Institut f{\"u}r Polymerforschung, Ackermannweg 10, 55128 Mainz, Germany}
\author{Frank H.L. Koppens} \affiliation{ICFO - Institut de Ci\`{e}ncies Fot\`{o}niques, The Barcelona Institute of Science and Technology, Castelldefels (Barcelona) 08860, Spain} \alsoaffiliation{ICREA - Instituci\'o Catalana de Re\c{c}erca i Estudis Avancats, 08010 Barcelona, Spain}
\author{Dmitry Turchinovich} \affiliation{Fakult{\"a}t f{\"u}r Physik, Universit{\"a}t Bielefeld, Universit{\"a}tsstr. 25, 33615 Bielefeld, Germany}
\author{Michael Gensch} \affiliation{Institute of Optical Sensor Systems,DLR, Rutherfordstr. 2, 12489 Berlin, Germany} \alsoaffiliation{Institut f{\"u}r Optik und Atomare Physik, Technische Universit{\"a}t Berlin, Strasse des 17. Juni 135, 10623 Berlin, Germany} 
\author{Sergey Kovalev} \email{s.kovalev@hzdr.de} \affiliation{Helmholtz-Zentrum Dresden-Rossendorf, Dresden, Germany}
\author{Klaas-Jan Tielrooij} \email{klaas.tielrooij@icn2.cat} \affiliation{Catalan Institute of Nanoscience and Nanotechnology (ICN2), BIST \& CSIC, Campus UAB, Bellaterra 08193, Barcelona, Spain}


\begin{document}
\setcounter{tocdepth}{1} 
\setcounter{page}{1}
\newpage

\begin{abstract}

Nonlinear optics is an increasingly important field for scientific and technological applications, owing to its relevance and potential for optical and optoelectronic technologies. Currently, there is an active search for suitable nonlinear material systems with efficient conversion and small material footprint. Ideally, the material system should allow for chip-integration and room-temperature operation. Two-dimensional materials are highly interesting in this regard. Particularly promising is graphene, which has demonstrated an exceptionally large nonlinearity in the terahertz regime. Yet, the light-matter interaction length in two-dimensional materials is inherently minimal, thus limiting the overall nonlinear-optical conversion efficiency. Here we overcome this challenge using a metamaterial platform that combines graphene with a photonic grating structure providing field enhancement. We measure terahertz third-harmonic generation in this metamaterial and obtain an effective third-order nonlinear susceptibility with a magnitude as large as 3$\cdot$10$^{-8}$m$^2$/V$^2$, or 21 esu, for a fundamental frequency of 0.7 THz. This nonlinearity is 50 times larger than what we obtain for graphene without grating. Such an enhancement corresponds to third-harmonic signal with an intensity that is three orders of magnitude larger due to the grating. Moreover, we demonstrate a field conversion efficiency for the third harmonic of up to $\sim$1\% using a moderate field strength of $\sim$30 kV/cm. Finally we show that harmonics beyond the third are enhanced even more strongly, allowing us to observe signatures of up to the 9$^{\rm th}$ harmonic. Grating-graphene metamaterials thus constitute an outstanding platform for commercially viable, CMOS compatible, room temperature, chip-integrated, THz nonlinear conversion applications.

\end{abstract}

\maketitle

\clearpage

The ability to perform nonlinear optical conversion has relevance for a broad range of current and future technologies, including ultrashort pulse generation, advanced spectroscopy, optical information processing and storage, telecommunications, light harvesting, bio-imaging, integrated optics, and quantum technologies \cite{Reshef2019,Kauranen2012,Miller2010,Maier2014}. For many of these applications it is crucial to find nonlinear optical material systems that allow for integration into existing nanophotonic platforms, and for operation with low power consumption. Thus, ideal nonlinear optical materials should be compatible with complementary metal-oxide–semiconductor (CMOS) technology, have a small material footprint, and have a large nonlinear susceptibility, such that large nonlinear effects occur for small incident optical fields. Furthermore, they should ideally operate at room temperature. Steady progress has been made in recent years, with suitable material systems operating in the visible and near-infared parts of the electromagnetic spectrum, for example using indium tin oxide with a thickness of tens of nanometers \cite{Alam2016,Alam2018}. However, currently no commercially viable nonlinear material system is available for incident light in the terahertz region (roughly 0.3 -- 30 THz) of the electromagnetic spectrum. This technological gap needs to be filled, because the THz region holds great promise for both scientific and technological applications, such as imaging, industrial quality control, gas sensing and wireless communications. All of the above would benefit tremendously from the availability of integrated nonlinear optical devices in the THz. 
\\

It is particularly appealing to use nonlinear optical materials that are thinner than the wavelength of light. In this case, different optical waves propagating through the material automatically remain in phase, thus removing the need for complex engineering to achieve phase matching. As a result, nonlinear optical conversion in ultimately thin, two-dimensional materials has received ample attention, as reviewed recently in Ref.\ \cite{You2019}. However, the challenge associated with using sub-wavelength materials is that -- even with extraordinary large nonlinear coefficients -- there is simply not much material to interact with, thus usually leading to low nonlinear conversion efficiencies. A promising strategy to overcome this challenge had already been identified before the advent of two-dimensional materials. Namely, photonic structures can be used to enhance light-matter interactions and thereby increase the nonlinear conversion efficiency. Typical photonic structures that have been used include nanoparticles \cite{Nie1997}, hole arrays \cite{Fan2006,Nieuwstadt2006}, nanocavity gratings \cite{Kats2010}, nanocavity-antenna systems \cite{Bailyn2011} split-ring resonators \cite{Liu2012,Fan2013}, and nanoantennas \cite{Kim2008, Maier2014, Alam2018}. Combining a photonic structure with a nonlinear material has, for example, made it possible to obtain a surface-enhanced Raman spectrum of a single molecule \cite{Nie1997} and to observe third-harmonic generation from a single semiconductor nanoparticle \cite{Maier2014}. 
\\

Interestingly, nonlinear effects in two-dimensional materials are particularly strong for light in the THz region. Indeed, two-dimensional, gapless Dirac materials, where electrons obey a linear energy-momentum dispersion relation, have shown very large nonlinear coefficients in the THz range. In particular, THz third-harmonic generation has been observed using monolayer graphene \cite{Hafez2018}, using the Dirac surface states in the topological insulator Bi$_2$Se$_3$ \cite{Giorgianni2016, Kovalev2020a}, and using the Dirac semimetal Cd$_3$As$_2$ \cite{Cheng2020, Kovalev2020}.The THz third-order nonlinear susceptibility of graphene was found to be especially large: $\vert \chi^{(3)} \vert \sim$10$^{-9}$ m$^2$/V$^2$ \cite{Hafez2018}. This value is many orders of magnitude larger than its $\chi^{(3)}$ in the near-infrared range, which is below $\sim$10$^{-16}$ m$^2$/V$^2$ \cite{Soavi2018}. The main reason for the strong THz nonlinearity is that it is the result of a different nonlinearity mechanism than the one in the near-infrared. Namely, the nonlinearity originates from the collective thermodynamic response of the quasi-free carriers in graphene. Their response to an incident THz pulse can be outlined as follows (see also Refs.\ \cite{Mics2015,Hafez2018, Hafez2020}). As a first step, graphene conduction-band electrons absorb the energy from incident THz light through Drude absorption. This energy is then distributed among all electrons leading to a thermal quasi-equilibrium within the electronic system, as it has not yet thermalized with the lattice. This is because the THz oscillation period and the electron cooling time are on the order of a picosecond, and thus much longer than the electron thermalization time, which is below 100 fs. The electronic temperature increase of the quasi-equilibrium Fermi-Dirac distribution is substantial due to the small electronic heat capacity of graphene \cite{Jensen2014}. Since the THz conductivity -- and therefore THz absorption -- decreases for increasing electron temperature (and Fermi energy above $\sim$0.1 eV) \cite{Tielrooij2013,Mics2015,Tomadin2018}, this gives rise to a large, ultrafast thermodynamic THz nonlinearity. Quasi-monochromatic THz excitation leads to ultrafast modulation of the graphene THz conductivity, and hence of the driven THz currents, within each cycle of driving THz field due to the electronic heating--cooling dynamics. This, in turn, leads to a nonlinear response that manifests as harmonic generation at odd-order multiples of the frequency of the driving field.  Given this strong nonlinearity and the fact that graphene can be integrated with CMOS-based chips, as demonstrated in Ref.\  \cite{Goossens2017}, it is a highly promising candidate material for commercially viable, chip-integrated, THz nonlinear conversion applications. One important hurdle, however, that needs to be overcome is to ensure sufficient light-matter interaction between incident THz light and a monolayer thin material. 
\\

In this work, we demonstrate a metamaterial system consisting of graphene and a metallic grating, where the grating provides field enhancement, thus leading to a strong increase of THz nonlinear light-matter interaction. We examine the generation of odd THz harmonics and observe that the third-harmonic intensity can be enhanced by more than three orders of magnitude due to the integration of the metal grating. We discuss quantitatively how this enhancement is achieved and what the limitations of this approach are, and the perspectives for further enhancing THz nonlinearities with this photonic approach. Finally, we show a field conversion efficiency of 1\% and signatures of higher harmonics, including the 9$^{\rm th}$ harmonic, using a moderate incident field strength of $\sim$30 kV/cm.
\\ 

\section{Results and discussion}

\subsubsection{Grating-enhanced THz third-order nonlinearity}

\begin{figure} [h!]
   \centering
 \includegraphics [scale=0.24]
   {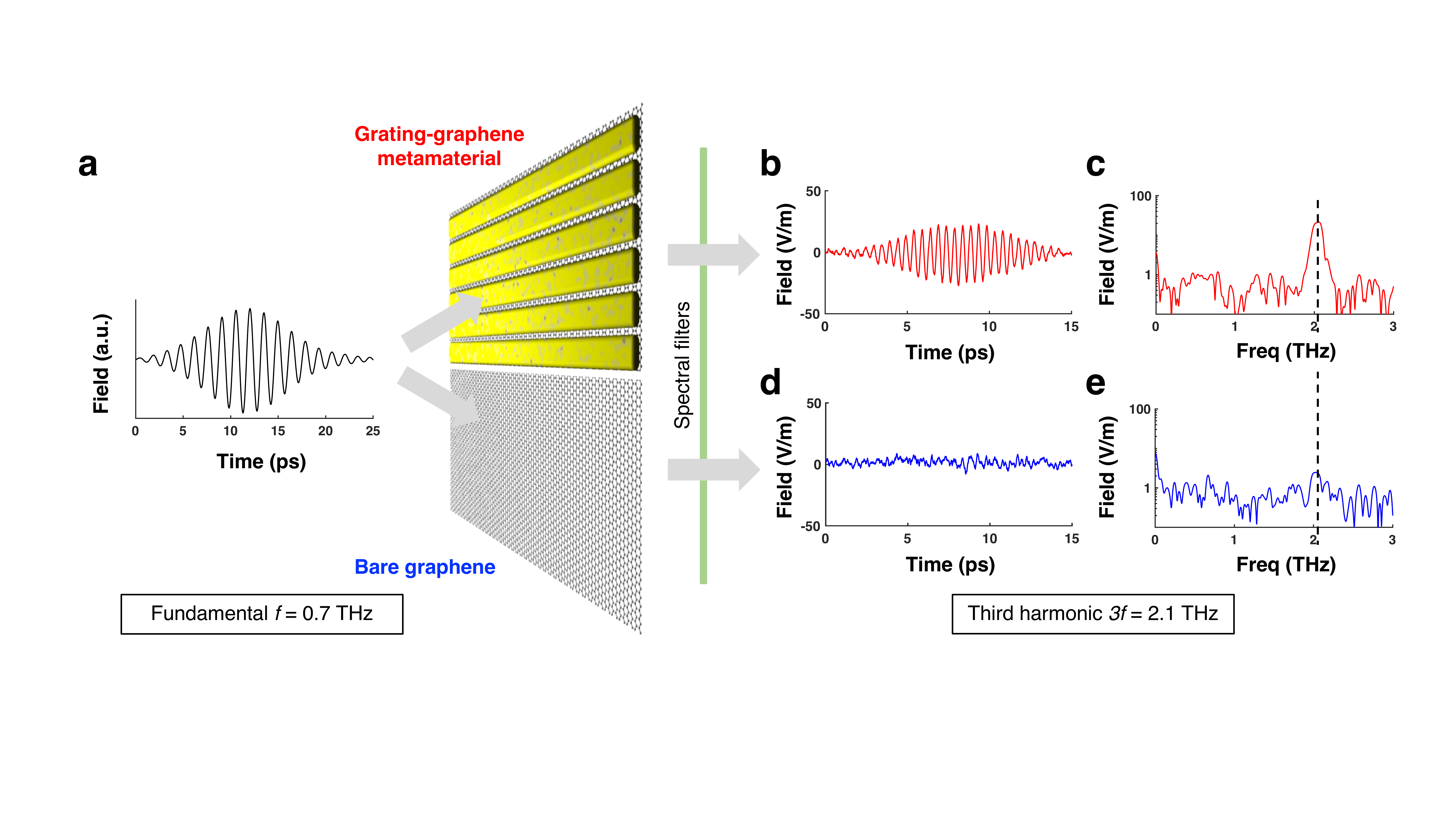}
   \caption{ \textbf{Enhanced third-harmonic generation in grating-graphene metamaterial.}
\textbf{a)} Schematic representation of the measurement configuration, with a multicycle THz waveform with fundamental frequency $f$ incident on a grating-graphene metamaterial sample (top) or a bare graphene sample (bottom). In the metamaterial, field enhancement occurs inside the metal gap, which leads to nonlinearly enhanced generation of third-harmonic signal. \textbf{b-c)} Measured THz field strength in the time \textbf{(b)} and frequency \textbf{(c)} domain for the grating-graphene metamaterial, with an incident field strength of 12 kV/cm. \textbf{d-e)} Measured THz field strength in the time \textbf{(d)} and frequency \textbf{(e)} domain for the bare graphene sample with an incident field strength of 13.6 kV/cm. Clearly, significantly more intense harmonic signal is created in the graphene metamaterial sample.
}
\label{Fig1a}
  \end{figure}

A natural and proven method to study the nonlinear properties of a material is to quantify the intensity or field strength of generated harmonic signal compared to the intensity or field strength of incident fundamental light. This method has already been applied successfully to graphene \cite{Hafez2018}, and therefore we follow the same experimental procedure. Briefly (see Methods for details), this method consists of focusing a narrowband THz waveform, generated by the accelerator-based TELBE beam facility \cite{Green2016}, onto a sample, and measuring the electric field of the transmitted waveform as a function of time through electro-optic sampling. In Fig.\ \ref{Fig1a}a we show a measured incident THz waveform with fundamental frequency $f$ = 0.7 THz in the time domain, together with a schematic of a grating-graphene metamaterial sample and a bare graphene sample. Figure \ref{Fig1a}b shows the signal in the time domain after passing through the metamaterial sample and spectral filters (2 bandpass filters at 2.1 THz) that suppress spectral components at the fundamental frequency. Clearly, the measured oscillations are faster than those of the incident light. The Fourier transform (see Methods and \ref{SuppFig12}) in Fig.\ \ref{Fig1a}c shows that these fast oscillations correspond to the generated third harmonic at $3f$ = 2.1 THz. 
\\

We first compare the nonlinear conversion efficiency of two samples, both consisting of wet-transferred monolayer graphene on quartz with an estimated Fermi energy of $\sim$0.23 eV (see Methods and \ref{SuppFig1} for details). Sample A has a gold metallic grating on top of the graphene, separated by a 2 nm thick layer of Al$_2$O$_3$. The metal grating contains metallic stripes with a metal width of $w_{\rm metal}$ = 18 $\mu$m, separated by gaps with a width of $w_{\rm gap}$ = 2 $\mu$m. This corresponds to a duty cycle $\eta = \frac{w_{\rm metal}}{w_{\rm metal} + w_{\rm gap}}$ of 90\%. Thus, 90\% of the graphene area is covered by metal and there is 10\% active graphene area. Sample B also contains 2 nm of Al$_2$O$_3$ on top of graphene, while not containing a metal grating. As a first indication that our approach of using a grating-graphene metamaterial to enhance nonlinear conversion works, we use a small incident field strength of $E_{\rm f}$ = 12 kV/cm, and observe clear third-harmonic signal for the grating-graphene metamaterial Sample A (Fig.\ \ref{Fig1a}b-c). However, for the bare graphene Sample B with a slightly larger incident field strength of 14 kV/cm there is no clearly observable third-harmonic signal in the time trace (see Fig.\ \ref{Fig1a}d). In the Fourier spectrum (see Fig.\ \ref{Fig1a}e) we observe a small signal just above the noise floor. Clearly, the metamaterial sample has a much stronger nonlinearity, even though only 10\% of the graphene is part of the active area. 
\\

\begin{figure} [h!]
   \centering
 \includegraphics [scale=0.68]
   {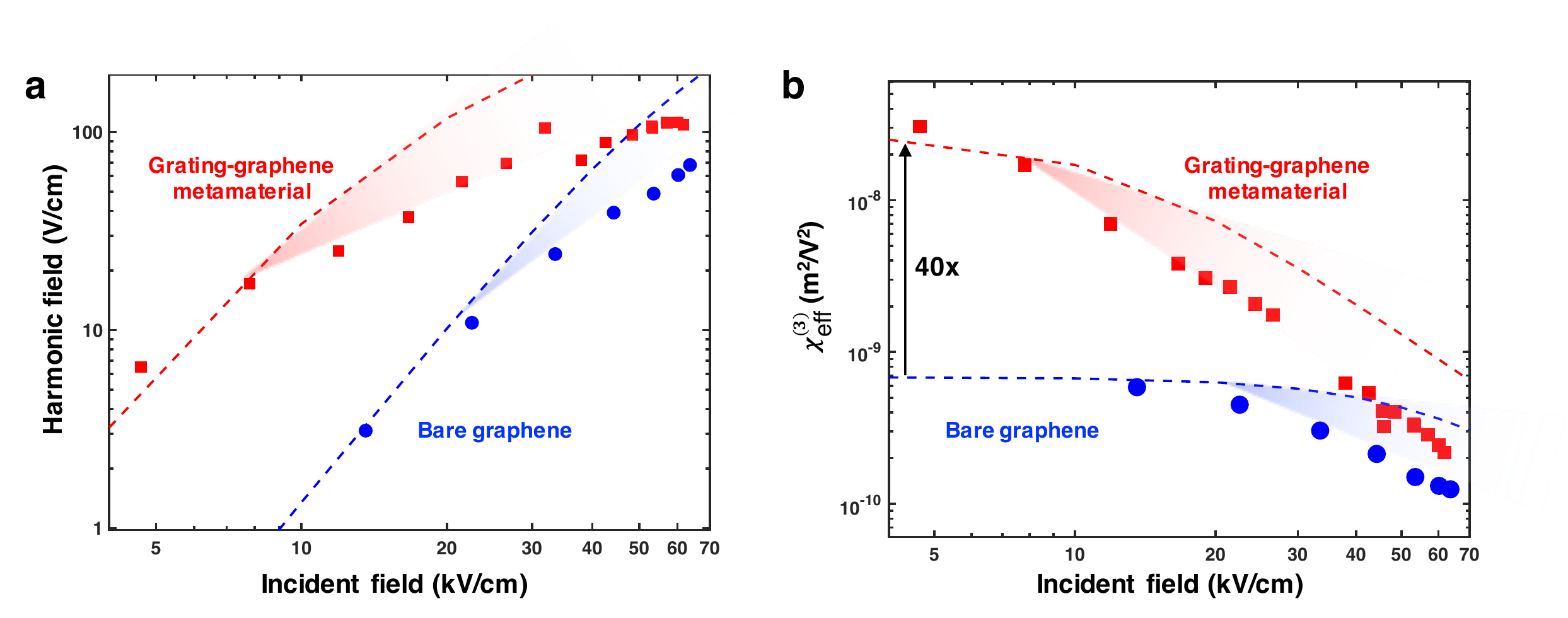}
   \caption{ \textbf{Nonlinearity of grating-graphene metamaterial \textit{vs.}\ bare graphene.}
\textbf{a)}
Comparison of third-harmonic intensity for grating-graphene metamaterial Sample A (red squares) and bare graphene Sample B (blue circles), as a function of peak field strength of the incident THz light. The blue dashed line is the result of split-step simulations based on thermodynamic nonlinearity without field enhancement. The red dashed line represents the same simulation, now with field-enhancement factor $M$ = 5 and active area $(1-\eta)$ = 10\%. 
\textbf{b)}
The extracted third-order nonlinear susceptibility $\chi^{(3)}_{\rm eff}$ as a function of incident peak field strength for grating-graphene metamaterial Sample A (red squares) and bare graphene Sample B (blue circles). The dashed lines show the results of the thermodynamic model. The decreasing trends with incident field strength indicate the occurrence of saturation effects. For low field strengths, the metal grating leads to a factor 40 -- 50 increase in $\chi^{(3)}_{\rm eff}$, corresponding to more than 3 orders of magnitude enhanced harmonic intensity. The red and blue shaded regions in both panels indicate the onset of saturation effects not captured by the model.
}
\label{Fig1b}
  \end{figure}

In order to quantify the nonlinear properties of our grating-graphene metamaterial, we compare how the field strength of the third-harmonic signal $E_{\rm 3f}$ scales with the incident field strength $E_{\rm f}$ (see Fig.\ \ref{Fig1b}a). These data points are obtained by taking the Fourier transform of transmitted time traces as in Fig.\ \ref{Fig1a}, and taking the peak of the power spectrum around the frequency of the third-harmonic peak at 2.1 THz. We observe that, for the lowest field strengths, $E_{\rm 3f}$ is more than an order of magnitude higher in the metamaterial sample than in bare graphene, whereas the third-harmonic signals are more similar for the two samples at the highest field strengths. We extract the third-order susceptibility using \cite{Hafez2018} 

\begin{equation}
\vert \chi_{\rm eff}^{(3)}\vert = \frac{E_{\rm 3f}}{E_{\rm f}^3}\frac{4 c\cdot n_{\rm 3f}}{3\pi f d} \hspace{0.2cm},
\end{equation} 

\noindent where $c$ is the speed of light, and $d =$ 0.3 nm is the thickness of graphene. For the refractive index of graphene at the third-harmonic frequency we conservatively use  $n_{\rm 3f}$ = 10, as in Ref.\ \cite{Hafez2018}. This value can be up to an order of magnitude larger, which would make the nonlinear coefficient an order of magnitude larger. The results in Fig.\ \ref{Fig1b}b show that for low incident field strength the grating-graphene metamaterial reaches an experimental value of $\vert \chi_{\rm eff}^{(3)}\vert$ = 3$\cdot$10$^{-8}$ m$^2$/V$^2$ (using $n_{\rm 3f}$ = 10). In ESU units, this corresponds to \cite{Ganeev2004} $\vert \chi_{\rm eff}^{(3)}\vert {\rm [ESU]} = \frac{9\cdot10^9}{4\pi}\vert\chi^{(3)}_{\rm eff}\vert = 21$ esu.  Expressed as a surface nonlinearity, we obtain $\vert \chi_{\rm 2D}^{(3)} \vert = \vert \chi_{\rm eff}^{(3)} \vert \cdot d$ = 10$^{-17}$m$^3$/V$^2$. We can also express the nonlinearity as the nonlinear refractive index $n_2$, where the intensity-dependent refractive index is given as \cite{Reshef2019} $n = n_0 + n_2 \cdot I_{\rm THz}$. Here $n_0$ is the linear refractive index and $I_{\rm THz}$ is the intensity of the incoming THz light. We estimate an $n_2$ well above 100 cm$^2$/GW (see Methods). Finally we note that the sign of the THz nonlinearity, and therefore of $\chi_{\rm eff}^{(3)}$ and $n_2$, is negative, since the refractive index decreases upon THz excitation (see also Ref.\ \cite{Hafez2018}), in agreement with the sign of the nonlinearity in the near-infrared \cite{Vermeulen2018}.
\\

Comparing the nonlinear susceptibility of the grating-graphene metamaterial Sample A with graphene without grating (Sample B), we find a value that is $\sim$50 times larger. The obtained value of $\vert \chi_{\rm eff}^{(3)} \vert$ is also more than three orders of magnitude larger than the highest THz nonlinearities reported for quantum well systems \cite{Sirtori1992,Markelz1996,Kono1997}. We can also compare the THz nonlinearity of our grating-graphene metamaterial with the near-infared nonlinearities found very recently for thin films of indium tin oxide (ITO) in a special frequency region with epsilon near zero, with a nonlinear refractive index $n_2$ of $\sim$0.1 cm$^2$/GW without field enhancement \cite{Alam2016} and $\sim$4 cm$^2$/GW with nanoantenna-induced enhancement \cite{Alam2018}. These nonlinear refractive indices are lower than what we find for our grating-graphene metamaterial in the THz regime. Overall, we can conclude that the THz nonlinearity of our grating-graphene metamaterial is remarkably large, compared to any nonlinear optical (meta)material at any wavelength, operating at room temperature. 
\\

We can understand the enhancement of the nonlinearity induced by the metal grating quantitatively by incorporating field enhancement into the thermodynamic model of Ref.\ \cite{Hafez2018, Mics2015}. For bare graphene Sample B, we straightforwardly use the incident field strength $E_{\rm f}$, the modal carrier density obtained from Raman measurements (see \ref{SuppFig1}), an estimated momentum scattering time, and standard picosecond cooling dynamics as the input parameters for the simulation. This produces the blue dashed lines in Fig.\ \ref{Fig1b}, which describe the experimental data quite well up to an incident field strength of $\sim$20 kV/cm. For higher field strengths, there is more saturation in the experimental data than what is predicted by the model. We will discuss saturation effects in more detail in the next section. For the simulations of metamaterial Sample A we use the exact same input parameters as for Sample B, with the difference that we use as incident field strength $M \cdot E_{\rm f}$, where $M$ is the field-enhancement factor. We estimate an average field-enhancement factor $M$ of $\sim$5 inside the gap of our grating, using rigorous coupled wave analysis (RCWA), following Ref.\ \cite{Zanotto2016} (see \ref{SuppFig10} and \ref{SuppFig9}). The resulting third-harmonic signal would correspond to a sample where the whole graphene area experiences 5-fold field enhancement, whereas in our case the active area is only $1 - \eta$ = 10\%. Therefore, we multiply the simulation results with a factor $\sqrt{1 - \eta}$, where the square root comes from the fact that we consider field, rather than intensity. The results are the red dashed lines in Fig.\ \ref{Fig1b}, which match the data for the lowest field strengths. Thus, the experimental results are in agreement with a grating-induced field enhancement of a factor $\sim$5. 
\\

Importantly, the third-harmonic signal is increased in a nonlinear fashion by the field enhancement. Indeed, we find that for the lowest incident field strengths, the model predicts an increase of the third harmonic field strength by a factor $\sim$40, corresponding to an intensity enhancement factor of $>$1000. We can understand these numbers analytically, as for the $n^{\rm th}$ harmonic we expect

\begin{equation}
\frac{E_{\rm nf}^{\rm (grating)}}{E_{\rm nf}^{\rm (no \hspace{0.1cm} grating)}} = M^n \cdot \sqrt{1 - \eta} \hspace{0.3cm}.
\end{equation} 

For $n$ = 3, $M$ = 5 and $\eta$ = 90\%, the analytical model gives the same factor of $\sim$40 as the numerical model based on the thermodynamic nonlinearity. This factor is in good agreement with the experimental observation of a $\vert \chi^{(3)}_{\rm eff} \vert$ that is $\sim$50 times larger for the grating-graphene metamaterial Sample A, in comparison with the bare graphene Sample B. The small discrepancy could come from a larger field-enhancement factor $M$, or from the fact that neither the thermodynamic nor the analytical model take into account that the grating could also lead to more efficient out-coupling of harmonic light to the far-field, where it is detected. We note that in the graphene-filled gap region, and in the perturbative low-field regime, the third harmonic-intensity is enhanced by a factor $M^6$ (see Eq.\ 1), which is more than four orders of magnitude. 
\\

\subsubsection{Saturation of harmonic generation}

We now discuss the occurrence of saturation effects, which follow clearly from Fig.\ \ref{Fig1b}b, where without saturation one would obtain a flat line. We first discuss two saturation mechanisms that are captured by the thermodynamic model. The first reason for saturation is the occurrence of saturable absorption: if the carrier temperature increases, the graphene conductivity decreases and the absorption of THz light is reduced (see also \ref{SuppFig30}). Eventually, the carrier temperature can become so high that virtually no incident THz light is absorbed, as shown in Ref.\ \cite{Mics2015}. The second reason that saturation occurs is that -- with constant absorption -- the increase in carrier temperature of graphene scales sub-linearly with the intensity of incident light, which is the result of the electronic heat capacity scaling with temperature \cite{Graham2012}. This sub-linear behavior starts to play a role when the carrier temperature increase $\Delta$T is larger than the initial carrier, or lattice, temperature (300 K). We remind that it is the increase in carrier temperature that gives rise to the THz nonlinearity \cite{Mics2015,Hafez2018}. Thus, the sub-linear scaling of carrier temperature with light intensity gives rise to a less strong nonlinearity. 
\\

There are also saturation effects that are not captured by our thermodynamic model, thus leading to a discrepancy between experimental data points and the results of the thermodynamic model (see the blue and red shaded areas in Fig.\ \ref{Fig1b}). We first discard the occurrence of light-induced damage to the grating or the graphene, as we did not observe any non-reversible effects of the generated harmonic signal when increasing and decreasing incident field strength, and confirmed that there was no damage by inspection with an optical microscope. We note that the damage observed in Ref.\ \cite{Liu2012}, inside their split-ring resonator gap occurred for a peak field strength of $\sim$4 MV/cm, an order of magnitude larger than our highest local field inside the grating gap, which was $M\cdot60$ = 300 kV/cm. Having discarded damage, we consider saturation related to the carrier temperature of graphene that goes beyond the thermodynamic model. We point out that for the metamaterial sample the carrier temperature reaches a value above 2000 K already for an incident field strength of $\sim$20 kV/cm (see \ref{SuppFig30}). With heat accumulating in both the electronic and phononic systems, phonon bottleneck effects lead to a longer cooling time for the hot carrier system. We have simulated the effect of slower cooling, and find that a factor 10 increase in cooling time leads to almost a factor 10 decrease in harmonic generation (see \ref{SuppFig5}). Furthermore, it gives rise to a red-shift of the harmonic signal, which we indeed observe (see \ref{SuppFig2}). At these elevated carrier temperatures increased phonon scattering can furthermore decrease the carrier mobility, thus giving rise to even stronger saturable absorption. Therefore, we ascribe the strong saturation effects we observe in the grating-graphene metamaterial, even for field strengths on the order of 10 kV/cm, to the comparably high carrier temperatures that we reach due to field enhancement. 
\\

\subsubsection{Photonic enhancement}

\begin{figure} [h!]
   \centering
 \includegraphics [scale=0.26]
   {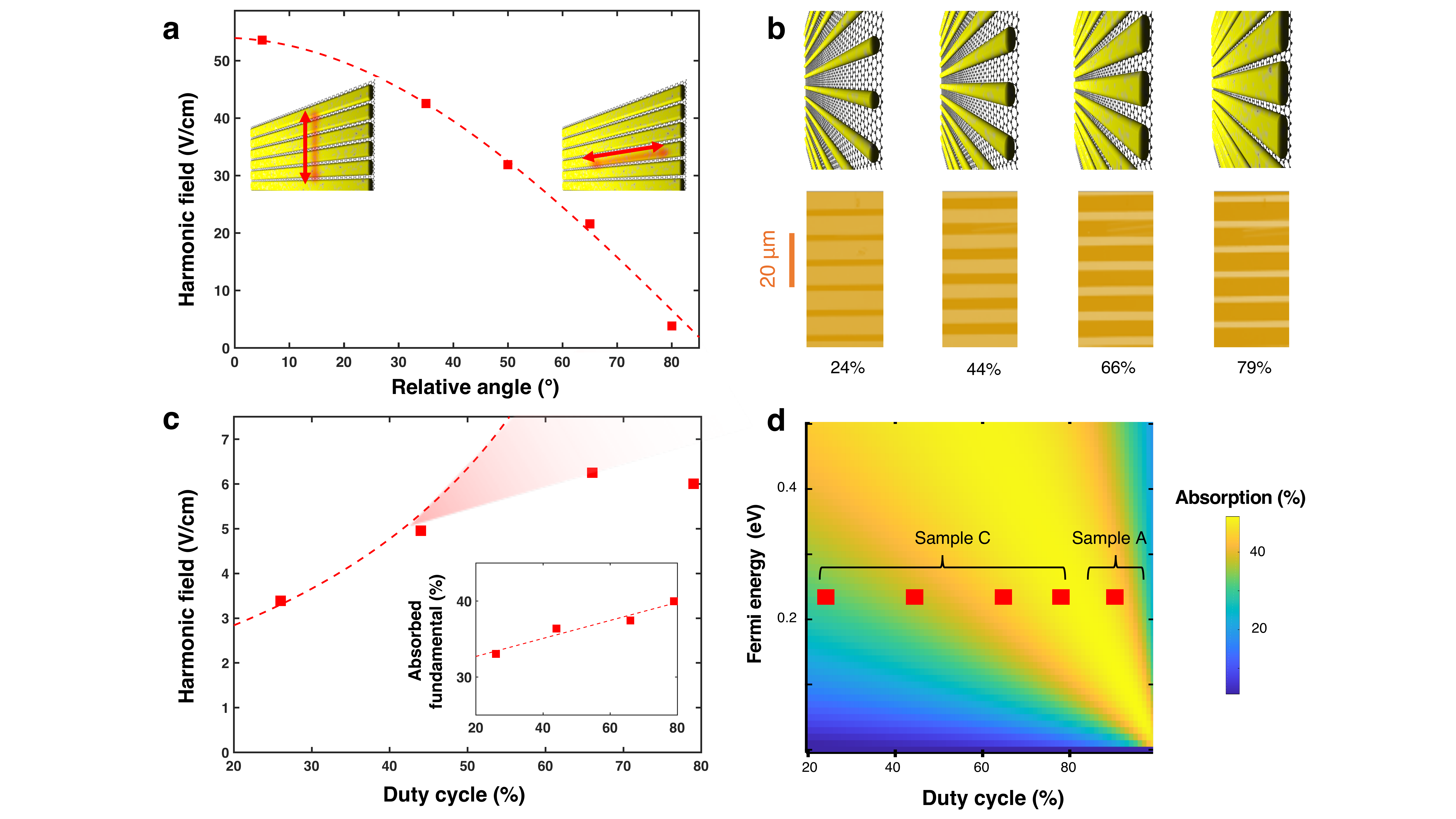}
   \caption{ \textbf{Effect of grating orientation and duty cycle. }
\textbf{a)} 
The field strength of the created third-harmonic signal as a function of orientation of the metal grating with respect to the polarization of the incoming fundamental THz light (red squares; Sample A; incident field strength of $\sim$32 kV/cm), as illustrated in the schematic insets. The dashed line is a $\sin^2$-function to guide the eye.
\textbf{b)}
Schematic representation (top) and optical images (bottom) of the four different grating-graphene areas of Sample C, with duty cycles of 24, 44, 66 and 79\%. Metal and gap widths are to scale. 
\textbf{c)}
The field strength of the created third-harmonic signal as a function of duty cycle using Sample C (red squares; incident peak field strength of $\sim$14 kV/cm), showing an increase, followed by saturation, for increasing duty cycle. The thermodynamic model (dashed line) matches the experimental results for the lowest duty cycles, before saturation effects occur, as indicated by the red shaded region. The inset shows the absorption of fundamental light \textit{vs.}\ duty cycle, showing larger absorption at higher duty cycle. This is obtained using (1 - transmitted fundamental intensity), and multiplying this by a constant, such that the area with a duty cycle of 24\% has a (1 - transmission) $\approx$ absorption of 33 \%, taken from RCWA simulations.  
\textbf{d)}
RCWA simulations of the absorption of grating-graphene structures as a function of Fermi energy and duty cycle, predicting higher absorption and thus higher field-enhancement factor $M$ for higher duty cycle. The increasing trend with Fermi energy at moderate duty cycles comes from the intrinsically larger THz absorption when there are more graphene charges to be accelerated. 
}
\label{Fig2}
  \end{figure}

We proceed with exploring how the intensity of the generated third-harmonic signal depends on the photonic enhancement of the metal grating, and what are the design considerations to optimize the grating. First, we verify the proper operation of the grating by varying its orientation with respect to the polarization of the incident field. The measurements in Fig.~1 were performed with the metal stripes perpendicular to the incoming polarization, where enhancement of field inside the gap of the grating structure is expected. Figure \ref{Fig2}a shows that by rotating the sample such that the orientation of the grating with respect to the incoming THz polarization changes over 90$^{\circ}$, the amount of generated harmonic signal is reduced to a value below our experimental noise floor. This is also as expected, because for parallel orientation of the polarization with respect to the metal stripes, no field enhancement occurs. Moreover, the grating in this case behaves as a THz filter that reduces THz transmission to the graphene film behind it. 
\\

In order to study how field enhancement depends on the duty cycle of the grating, we use a grating-graphene metamaterial sample -- Sample C -- which contains four areas with metal gratings with different duty cycles, varying from 24\% to 79\% (see Fig.\ \ref{Fig2}b). A larger duty cycle corresponds to a smaller metallic gap size, which should lead to larger field enhancement. Figure \ref{Fig2}c shows that with increasing duty cycle, the amount of generated harmonic indeed increases, although it saturates for the largest duty cycles. The reason that we don't see any further increase in detected third harmonic above $\eta =$ 70\% (see Fig.\ \ref{Fig2}c), is the occurrence of saturation effects, as discussed above. In order to gain further insights, we have performed RCWA simulations as a function of duty cycle and Fermi energy of graphene (see Fig.\ \ref{Fig2}d). The results show that the total amount of THz absorption increases with increasing duty cycle, before decreasing quickly upon approaching $\eta$ = 100\%. This total absorption is the sum of the absorption of the active graphene area inside the gap of the grating, and of the inactive area that is covered by metal. The field in the graphene region below the metal stripes is very small, so absorption there is negligible (see \ref{SuppFig10}a). For our samples with Fermi energy of $\sim$0.23 eV, we expect maximum absorption for a duty cycle of 80--90\%. This behavior can be understood from the trade-off between increasing field enhancement \textit{vs.}\ decreasing active graphene area with increasing duty cycle. We measured the transmitted fundamental intensity for Sample C with varying duty cycle, and use it to examine how the absorption $\approx$ (1 - transmitted fundamental) scales with duty cycle (see inset of Fig.\ \ref{Fig2}c). We confirm that the overall THz absorption (active plus inactive area) increases moderately with increasing duty cycle. Since the active area decreases with increasing duty cycle, this indicates increasingly strong field enhancement inside the gap. 
\\

Remarkably, the absorption simulations suggest that a grating-graphene metamaterial, despite the small active area, can reach almost 50\% absorption of all incident THz photons onto the metamaterial, for monolayer graphene with a standard mobility of 2000 cm$^2$/Vs. This is $\sim$3 times more than the amount of absorption in bare graphene with a Fermi energy of 0.23 eV, as estimated for Sample A (see \ref{SuppFig1}). With a duty cycle of $\eta$ = 90\%, \textit{i.e.}\ 10\% active area, the simulations give an overall absorption of 45\% in the metamaterial sample, which corresponds to an effective absorption of 450\% in the active graphene area. This is a photonic absorption enhancement of $\sim$26, which corresponds to a field enhancement of $M \sim$ 5 inside the gap, which is in good agreement with the electric field simulations in \ref{SuppFig10}. For the lowest duty cycle we studied experimentally, $\eta =$ 24\%, the overall THz absorption is still around 30\%. However the field enhancement $M$ inside the gap is only $\sim$1.5. Thus, in order to maximize the nonlinear susceptibility, the grating-graphene metamaterial would have a duty cycle approaching -- without reaching -- 100\%. The second parameter to optimize is the Fermi energy of graphene, where we note that a smaller Fermi energy means less intrinsic THz absorption. Interestingly, the simulations indicate that the combination of graphene with a Fermi energy approaching the Dirac point, and a metal grating with a duty cycle approaching 100\% still leads to absorption close to 50\%. As an outlook of what is possible with our nonlinear grating-graphene metamaterial platform, we calculate the expected nonlinearity for a grating with a duty cycle of 99\% and graphene with a Fermi energy of 0.1 eV. For such a grating, a period of 20 microns would require a gap size of 200 nm, which is possible using nanofabrication techniques, such as electron beam lithography or focused ion beam milling. We estimate that we can reach $M$ = 30 for $\eta =$ 99\% and $E_{\rm F}$ = 0.1 eV.  This would increase the third-harmonic intensity by a factor $M^6 \cdot (1 - \eta) \approx$ 10$^7$ and increase the third-order THz nonlinearity to above 10$^{-6}$m$^2$/V$^2$. 
\\

\subsubsection{Overall conversion efficiency and ninth harmonic generation}

\begin{figure} [h!]
   \centering
 \includegraphics [scale=0.6]
   {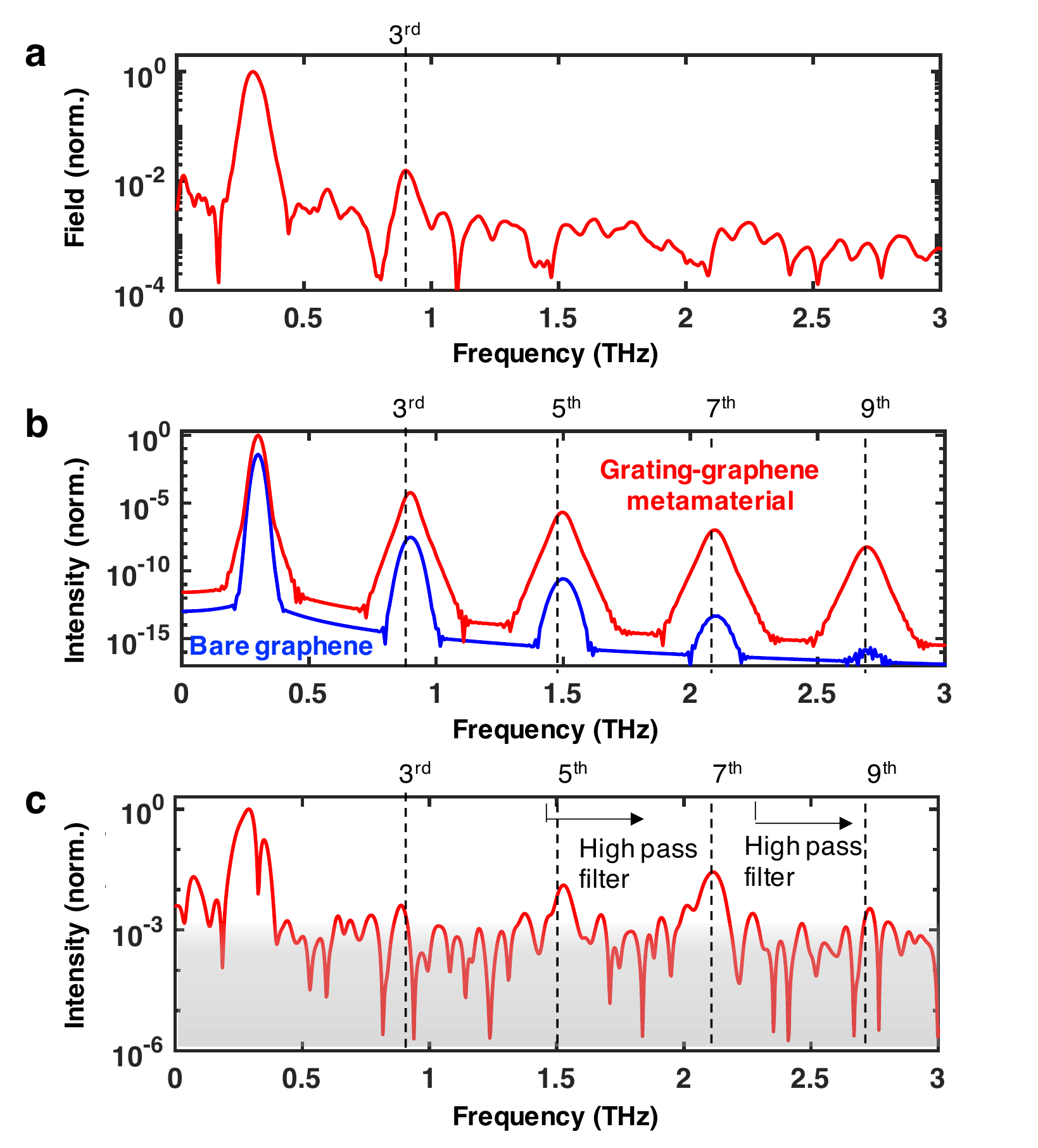}
   \caption{ \textbf{Conversion efficiency and 9$^{\rm th}$ harmonic.}
\textbf{a)} Spectral content of the THz waveform after grating-graphene metamaterial Sample A using a table-top source with 30 kV/cm at 0.3 THz. We did not use any spectral filters in the detection pulse, thus allowing to observe quantitatively the amount of transmitted fundamental and generated harmonic in the spectrum, giving a field conversion efficiency of $\sim$1\%. 
\textbf{b)}
Split-step simulation results of the effect of a grating-induced enhancement factor $M =$ 5 on the possibility to observe higher harmonics. The incident field is 30 kV/cm at 0.3 THz. The blue (red) line is without (with) field enhancement. At the fundamental frequency, the intensity is enhanced by  $M^2$ = 25, whereas due to the higher order nonlinear processes that give rise to higher harmonics, field confinement leads to even larger enhancement of harmonic intensity -- for example, around 7 orders of magnitude for the 9$^{\rm th}$ harmonic. 
\textbf{c)} 
Experimental results using our table-top source with incident field strength of $\sim$30 kV/cm and grating-graphene metamaterial Sample A. We put high-pass filters in the detection path, in order to focus on higher harmonics, thus suppressing strongly the fundamental and 3$^{\rm rd}$ harmonic. We observe very clear signatures of the 5$^{\rm th}$ and 7$^{\rm th}$ harmonic, and even the 9$^{\rm th}$ harmonic. The grey shaded area indicates the estimated experimental noise floor. 
}
\label{Fig4}
  \end{figure}

In addition to optimizing the nonlinear susceptibility, as we just described, an important goal towards applications is to obtain a large nonlinear conversion efficiency. We therefore test Sample A (with 90 \% duty cycle) using an experimental geometry where we directly obtain the conversion efficiency. To this end we use a table-top THz source producing a moderate peak field strength of $\sim$30 kV/cm. Briefly (see Methods for details), we use an amplified Ti:sapphire laser system incident on a LiNbO$_3$ nonlinear crystal, together with two 0.3 THz bandpass filters, to create a multicycle THz waveform in a narrow band around 0.3 THz. As a detection crystal we use a 100 $\mu$m thick ZnTe crystal with $<$110$>$ cut on top of 2 mm $<$100$>$ ZnTe \cite{Turchinovich2007}. This gives us a flat spectral response and eliminates reflections, meaning that the sensitivities for the fundamental at 0.3 THz and the third harmonic at 0.9 THz are identical. Crucially, we measure the transmitted THz waveform without any filters behind the sample, thus transmitting both fundamental and created harmonic signal. The Fourier transform is shown in Fig.\ \ref{Fig4}a and clearly contains both fundamental at 0.3 THz and third harmonic at 0.9 THz. We directly obtain the ratio between generated third-harmonic and transmitted fundamental fields, and find a remarkably large value of 1.6\%, which -- after correcting for an estimated $\sim$45\% absorption of THz fundamental in the metamaterial -- corresponds to a nonlinear conversion efficiency of $\sim$1\%. This is a highly encouraging, and -- to the best of our knowledge -- unprecedented value for the field conversion efficiency of THz third-harmonic generation using such a modest field strength ($\sim$30 kV/cm). 
\\

As a final experiment, we examine harmonics beyond the third harmonic. Since grating-induced field enhancement increases harmonic generation in a nonlinear way (see Eq.\ 1), this effect will be even more pronounced for higher harmonics. Analytically, we would expect the intensity of the $n$ = 9$^{\rm th}$ harmonic, for example, to be enhanced by 11 orders of magnitude using our grating with a duty cycle of 90\% and field-enhancement factor $M$ = 5. This, however, does not take any saturation effects into consideration. Therefore, we obtain a more accurate estimate using the thermodynamic split-step calculations of THz harmonic generation following Ref.\ \cite{Hafez2018}. We use a modest incident field strength of 30 kV/cm, in order to avoid the occurrence of saturation effects that are not captured by our model. The simulation results in Fig.\ \ref{Fig4}b predict that the 9$^{\rm th}$-harmonic intensity for our grating-graphene nonlinear metamaterial will be enhanced by more than 7 orders of magnitude compared to bare graphene. To verify this experimentally, we use our table-top setup, now with two high-pass filters behind Sample A and a GaP detection crystal, which allows for a larger acceptance THz spectral bandwidth, as compared to the previously used ZnTe crystal. Figure \ref{Fig4}c shows the obtained spectrum with very clear 5$^{\rm th}$ and 7$^{\rm th}$ harmonic. Remarkably, there is even a discernible signature of the 9$^{\rm th}$ harmonic just above our estimated noise floor. We remark that the observation of 9$^{\rm th}$ harmonic signal with incident light in the near-infrared in the strongly nonlinear epsilon-near zero material CdO required a much larger field strength, namely $>$1 MV/cm \cite{Yang2019}. This highlights the exceptional strength of the THz nonlinearity of our material system. 
\\ 

\section{Conclusion}

We have demonstrated a grating-graphene metamaterial, where grating-induced field enhancement signficantly increases the already large THz nonlinearity of graphene. For a grating with a duty cycle of 90\%, we have obtained an effective nonlinear susceptibility $\chi_{\rm eff}^{(3)}$ as high as -3$\cdot$10$^{-8}$ m$^2$/V$^2$, and a third-harmonic generation conversion efficiency of $\sim$1\% using a moderate incident field strength of $\sim$30 kV/cm. We qualitatively understand the occurrence of this large nonlinearity, as being the result of the incident field being enhanced inside the gaps of the grating. Due to the nonlinear nature of harmonic generation, the intensity of the third-harmonic signal is increased by a factor much larger than the field enhancement. Quantitatively, we find agreement using a field-enhancement factor of 5, as obtained from RCWA calculations, and calculating harmonic generation analytically in the perturbative regime, and numerically using a thermodyamic model of the THz nonlinearity beyond the perturbative regime. We have shown that for incident field strengths above $\sim$10 kV/cm saturation effects related to the high carrier temperature occur. Low incident field strengths are in any case most relevant, as commercial applications require low power consumption. Based on our understanding of the metal grating, we have shown perspectives for further improving the nonlinear conversion efficiency of the grating-graphene system, in particular by increasing the duty cycle and decreasing the Fermi energy. Finally, we have shown that the nonlinear enhancement allows for creating higher harmonics (up to the 9$^{\rm th}$). The grating-graphene metamaterial that we have introduced here relies on standard, CMOS-compatible fabrication technology using CVD-grown graphene, has a very small material footprint, and sufficiently high nonlinear conversion efficiency to ensure low power consumption. We thus conclude that it is an outstanding platform for commercially viable, chip-integrated, THz nonlinear conversion applications.
\\

\section{Methods}

\subsubsection{Sample fabrication}

All samples were prepared starting with Infrasil quartz substrates with excellent THz transparency. For Samples A and B, we then used standard PMMA-assisted wet transfer of CVD-grown graphene (Grolltex). This was followed by atomic layer deposition (ALD) of 2 nm of Al$_2$O$_3$. For Sample A, we then used optical lithography and thermal evaporation of 50 nm of gold with titanium as sticking layer. Samples A and B, which are used for comparing the grating-induced harmonic enhancement, were prepared concurrently, under identical conditions with the exact same source materials,  thus ensuring that their key properties -- Fermi energy and mobility -- are nearly identical. For Sample C, we first fabricated the metal grating, followed by ALD of 2 nm of Al$_2$O$_3$ and then standard PMMA-assisted wet transfer of CVD-grown graphene. See \ref{SuppFig1} for a schematic overview of the samples and Raman characterization of Sample B. We note that RCWA calculations show no significant difference between samples where the graphene is placed on top or below the metal grating (see \ref{SuppFig9}). 
\\

\subsubsection{Experimental setups}

Measurements were all performed at the TELBE facility at Helmholtz-Zentrum Dresden-Rossendorf, at room temperature and ambient conditions\cite{Green2016}. The super-radiant source was tuned to provide multicycle, monochromatic, linearly polarized THz pulses with a frequency of 0.7 THz (20\% FWHM bandwidth) at a repetition rate of 100 kHz. The field strength was tuned using a pair of wiregrid polarizers. We used two band-pass filters at 2.1 THz behind the sample to reduce transmitted fundamental. The transmitted THz waveform, containing mainly re-emitted third-harmonic signal, was detected with electro-optic sampling (EOS) in a 1.9 mm thick ZnTe nonlinear crystal using a synchronized Ti:sapphire laser system with 100 fs pulses \cite{Kovalev2017}. In the analysis of the transmitted signals, we took into account the properties of the two filters, as well as the wavelength-dependent EOS response of ZnTe. For more details, see Ref.\ \cite{Hafez2018}.
\\

For the measurements in Fig.\ \ref{Fig4}, we used a Ti:sapphire amplified laser system to pump a nonlinear LiNbO$_3$ crystal under tilted pulse front configuration \cite{Yeh2007}, to generate THz radiation with a broadband, linearly polarized THz waveform at a repetition rate of 1 kHz. We then used a band-pass filter at 0.3 THz to create a spectrally more narrow, multicycle THz waveform. We detected the transmitted THz radiation in Fig.\ \ref{Fig4}a, without passing through any filters, using a 100 $\mu$m thick ZnTe crystal with $<$110$>$ cut on top of 2 mm $<$100$>$ ZnTe \cite{Turchinovich2007}. For the data in Fig.\ \ref{Fig4}c we performed EOS using a 300 $\mu$m thick GaP crystal and we added two high-pass filters with cut-off frequencies at 1.4 THz and 2.2 THz, in order to resolve higher harmonics. 
\\

\subsubsection{Calculation of nonlinear refractive index}

For Kerr-like optical nonlinearities the nonlinear refractive index is given by \cite{Reshef2019}

\begin{equation}
n_2 = \frac{3\chi_{\rm eff}^{(3)}}{4 n_{\rm 3f}^2\epsilon_0c}  \hspace{1cm} {\rm for} \hspace{1cm}{\vert n_2\vert  I_{\rm THz}}/{n_0} \ll 1 \hspace{0.2cm},
\end{equation}

\noindent Here $\epsilon_0$ is the vacuum permittivity. Using our $\chi_{\rm eff}^{(3)}$ of -3$\cdot$10$^{-8}$ m$^2$/V$^2$ and $n_{\rm 3f}$ = 10, we obtain an $n_2$ close to -10$^6$ cm$^2$/GW. This number is valid for very small incident field strength. We note that even with a field strength of 5 kV/cm (equivalent to an intensity of $I_{\rm THz} = c \epsilon_0 n \vert E \vert ^2/2 \approx$ 10$^{-5}$ GW/cm$^2$), we do not meet the criterion above. Therefore, we make an alternative estimate of $n_2$ that is valid for the experimentally used field strengths. We consider a field strength of $\sim$50 kV/cm, where the third-harmonic signal is strongly saturated, indicating that the carrier temperature is so high that the THz conductivity is close to zero (see also Refs.\ \cite{Hafez2018, Mics2015}). This means that the refractive index has decreased from $\sim$10 to something close to 2.5 - the refractive index of graphite\cite{Blake2007}. We thus use $n_2 = (n - n_0) / I_{\rm THz}$, with $(n - n_0)$ = -7.5 and $I_{\rm THz}$ = 0.025 GW/cm$^2$ (for a field strength of 50 kV/cm), obtaining $n_2$ = -300 cm$^2$/GW. So the nonlinear refractive index for our experimental conditions is $\vert n_2 \vert >$100 cm$^2$/GW. 
\\

\section{Acknowledgments}

We thank David Saleta and Jake Mehew for Raman chacaracterization and analysis, Z. Wang for assistance in the HHG experiments, and Ulf Lehnert and Jochen Teichert for technical assistance. K.-J.T. acknowledges funding from the European Union’s Horizon 2020 research and innovation program under Grant Agreement No. 804349 (ERC StG CUHL) and financial support through the MAINZ Visiting Professorship. ICN2 was supported by the Severo Ochoa program from Spanish MINECO Grant No. SEV-2017-0706. Parts of this research were carried out at ELBE at the Helmholtz-Zentrum Dresden-Rossendorf e.V., a member of the Helmholtz Association. N.A., S.K., and I.I. acknowledge support from the European Union’s Horizon 2020 research and innovation program under grant agreement No. 737038 (TRANSPIRE). X.J. acknowledges the support from the Max Planck Graduate Center with the Johannes Gutenberg-Universität Mainz (MPGC).

\bibliography{GratingBib_12}

\clearpage

\clearpage

\clearpage

\renewcommand\thefigure{\textbf{Supp.\ Fig.\ \arabic{figure}}}
\setcounter{figure}{0}
\renewcommand{\figurename}{}

\clearpage

\textbf{SUPPLEMENTARY FIGURES}

\begin{figure} [h!]
   \centering
 \includegraphics [scale=0.9]
   {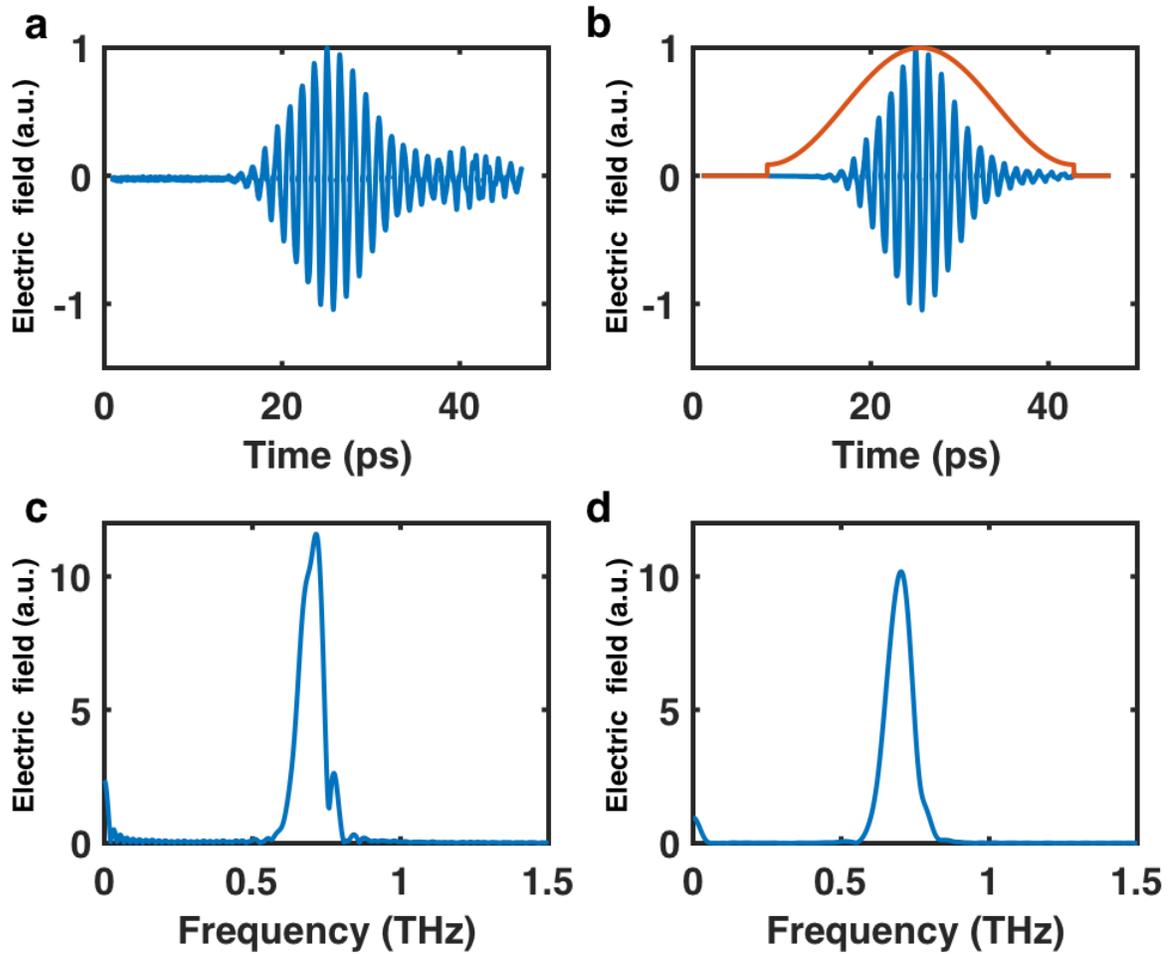}
   \caption{ \textbf{Data analysis with window function}
\textbf{a)} A typical time trace (normalized) as created from the super-radiant source, measured through EOS. 
\textbf{b)} The same trace as in panel \textbf{a} in blue, multiplied by the Hamming window function in red. 
\textbf{c-d)} The Fourier transforms of the traces in panel \textbf{a} \textbf{(c)} and in panel \textbf{b} \textbf{(d)}. Multiplying with the window function removes artifacts in the Fourier spectrum. 
}
\label{SuppFig12}
  \end{figure}

\clearpage

\begin{figure} [h!]
   \centering
 \includegraphics [scale=0.7]
   {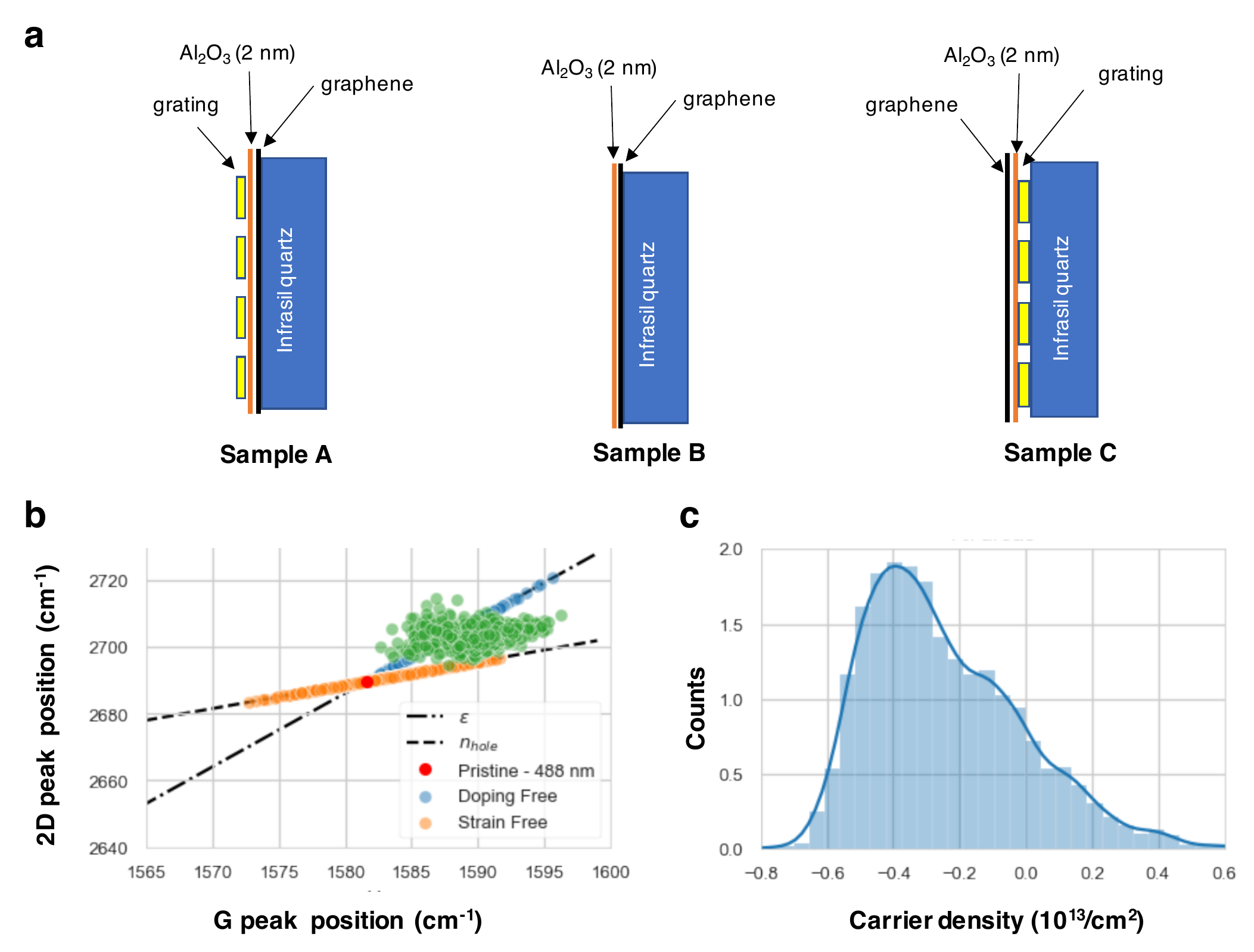}
   \caption{ \textbf{Sample description and characterization. }
\textbf{(a)} Schematic overview of the different sample geometries, where for Sample A, the graphene transfer was done before fabricating the metallic grating, for Sample B, no grating was fabricated, and for Sample C, the graphene transfer was done after the metallic grating.
\textbf{b)} Raman characterization of a sample from the same batch as Sample A and Sample B, without grating, plotting the position of the 2D peak \textit{vs.}\ the G peak. 
\textbf{c)} The strain-doping analysis yields a modal carrier density of 4$\cdot$10$^{12}$/cm$^2$, or $E_{\rm F}$ = 0.23 eV.
}
\label{SuppFig1}
  \end{figure}

\clearpage

\begin{figure} [h!]
   \centering
 \includegraphics [scale=0.65]
   {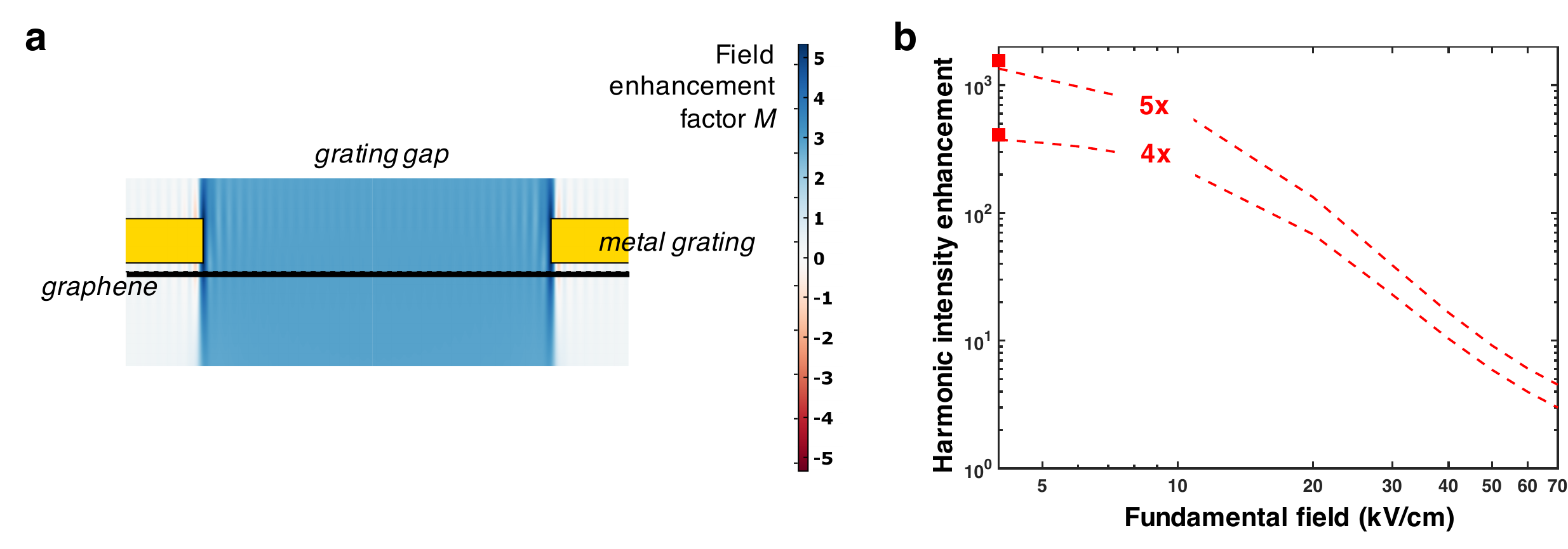}
   \caption{ \textbf{RCWA simulations of field enhancement.}
\textbf{a)} RCWA simulations, following Ref.\ \cite{Zanotto2016}, of the field enhancement for the geometry of Sample A, with grating on top of graphene and a duty cycle $\eta$ = 90\%. The field-enhancement factor $M$ according to this simulation is 4-5.
\textbf{b)} Enhancement factor of third-harmonic generation due to a field-enhancement factor $M$ = 4 and $M$ = 5. Dashed lines show simulation results using the thermodynamic model. The red squares show the analytically calculated harmonic intensity enhancement, according to $M^{2n}\cdot(1 - \eta)$, where $n$ = 3 for the third harmonic. This analytical result nicely matches the simulations for low field strength. The decrease in enhancement factor for higher field strength is due to saturation effects. 
}
\label{SuppFig10}
  \end{figure}

\clearpage

\begin{figure} [h!]
   \centering
 \includegraphics [scale=0.9]
   {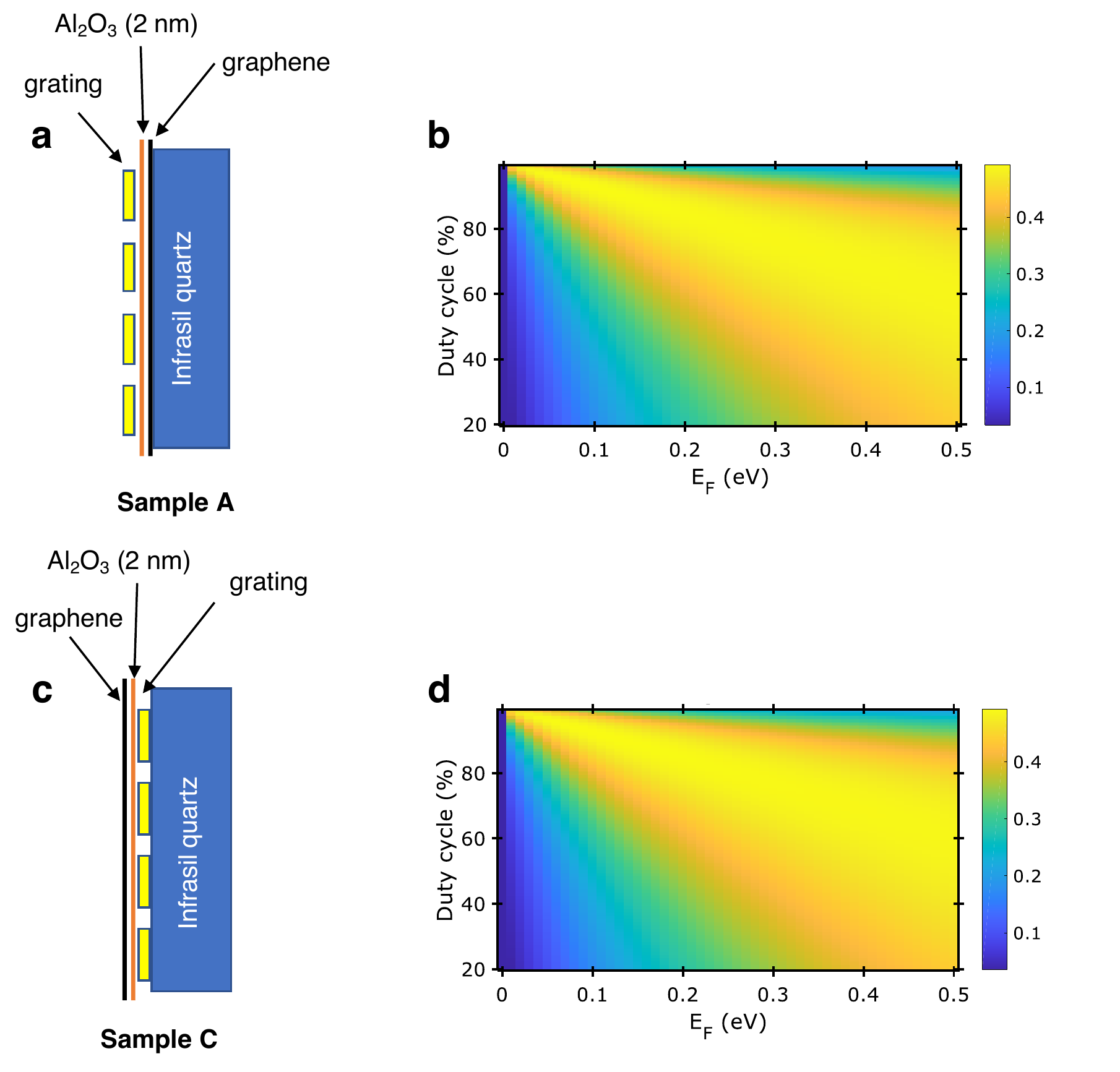}
   \caption{ \textbf{RCWA simulations of absorption enhancement.}
\textbf{a)} Sample geometry with metallic grating on top of graphene. This is the geometry of Sample A with duty cycle 90\%.
\textbf{b)} Simulated absorption for incident light at 0.7 THz on a grating-graphene metamaterial structure as in panel a. For graphene with $E_{\rm F}$ = 0.23 eV and mobility $\sim$2000 cm$^2$/Vs. The absorption reaches almost 50\%. Note that this is overall absorption in the entire area. Therefore, 50\% absorption for an 80\% duty cycle means 225\% absorption by the graphene in the gap. Since bare graphene with the same Fermi energy and mobility absorbs $\sim$17\%, absorption in the gap is enhanced by a factor $\sim$15, so the estimated field enhancement is $M \approx \sqrt{15} \approx$ 4. With 45\% absorption for a duty cycle of 90 \% (Sample A), these simulations suggest a field enhancement of $M \approx$ 5. 
\textbf{c)} Sample geometry with graphene on top of metallic grating. This is the geometry of Sample C with varying duty cycle. 
\textbf{d)} Simulated absorption under the same conditions as in panel b, now with the sample geometry as in panel c. The results are nearly identical. 
}
\label{SuppFig9}
  \end{figure}

\clearpage

\begin{figure} [h!]
   \centering
 \includegraphics [scale=0.6]
   {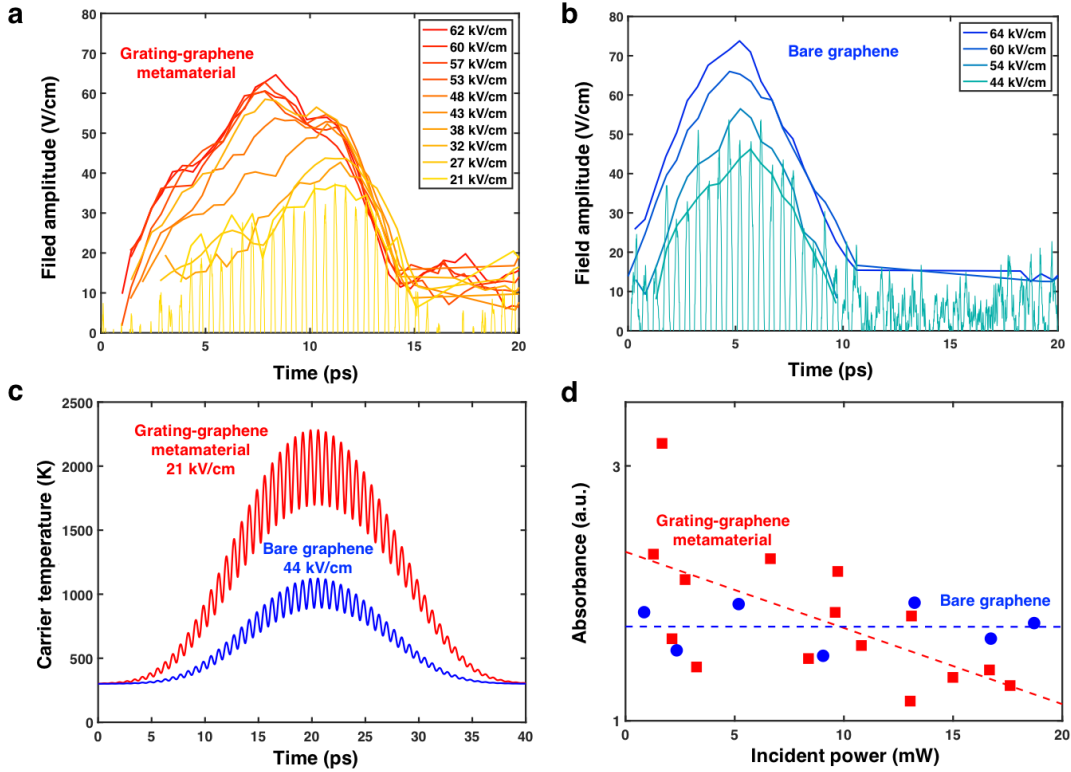}
   \caption{ \textbf{Saturation effects.}
\textbf{a-b)} The envelope of the transmitted THz waveforms as a function of time for different incident field strengths for grating-graphene metamaterial Sample A with 90\% duty cycle \textbf{(a)} and bare graphene Sample B \textbf{(b)}. This envelope is obtained using a peak finder algorithm. For the lowest field strengths, the time traces are also shown (thinner lines). These envelope functions show faster buildup and stronger saturation of the transmitted signal for the grating-graphene metamaterial sample. 
\textbf{c)} Thermodynamic split-step simulations of the temperature of the graphene electronic system upon interaction with incident THz light with 44 kV/cm without field enhancement (blue) and for 21 kV/cm with 5-fold field enhancement (red). For carrier temperatures above $\sim$600 K, we expect first saturation effects to occur, due to the sub-linear dependence of carrier temperature with incident power. Above $\sim$1500 K, additional saturation effects are expected, such as saturable absorption, slower carrier cooling and lower heating efficiency due to competition with relaxation to strongly coupled optical phonons. 
\textbf{d)} 
The absorbance (intensity of incident power divided by transmitted power) of fundamental light, as a function of incident power for graphene with (red squares, Sample A) and without (blue circles, Sample B) grating, together with linear fits to guide the eye (dashed lines). For bare graphene, the absorption is roughly constant (see dashed blue line). For the grating-graphene material sample, the absorbance clearly decreases with incident power (see red dashed line). This saturable absorption effect due to the high carrier temperature, is likely involved in the decrease of field conversion efficiency at higher peak field strengths. 
}
\label{SuppFig30}
\end{figure}

\clearpage

\begin{figure} [h!]
   \centering
 \includegraphics [scale=0.7]
   {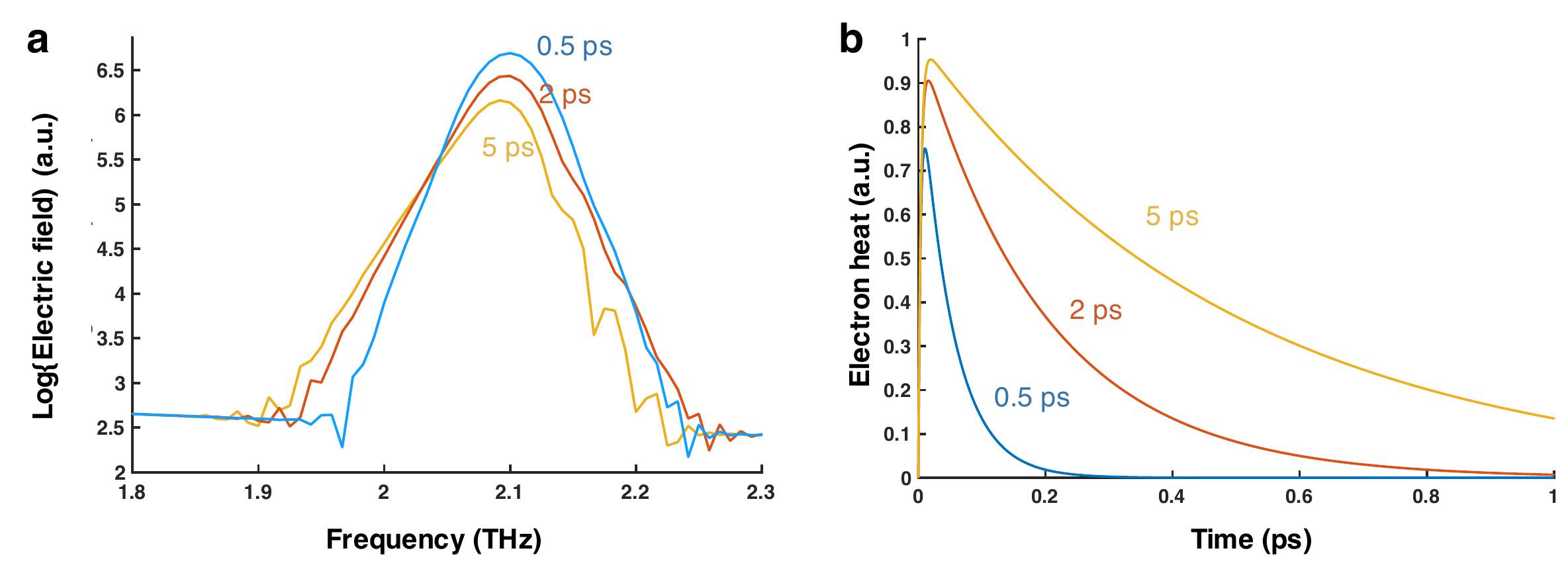}
   \caption{ \textbf{Effect of cooling dynamics.}
\textbf{a)} Thermodynamic split-step simulations of the created third harmonic for three different cooling times, with the heating-cooling dynamics as in panel \textbf{(b)}. When using a 10x longer cooling time, we find that a factor $\sim$10 less harmonic is generated. Furthermore, a red shoulder appears, just as the experimental results in \ref{SuppFig2} show. This suggests that a change in cooling dynamics, due to the high carrier temperature, likely plays a role in the strong saturation of third-harmonic signal for increasing incident field strength. 
}
\label{SuppFig5}
  \end{figure}

\clearpage

\begin{figure} [h!]
   \centering
 \includegraphics [scale=0.7]
   {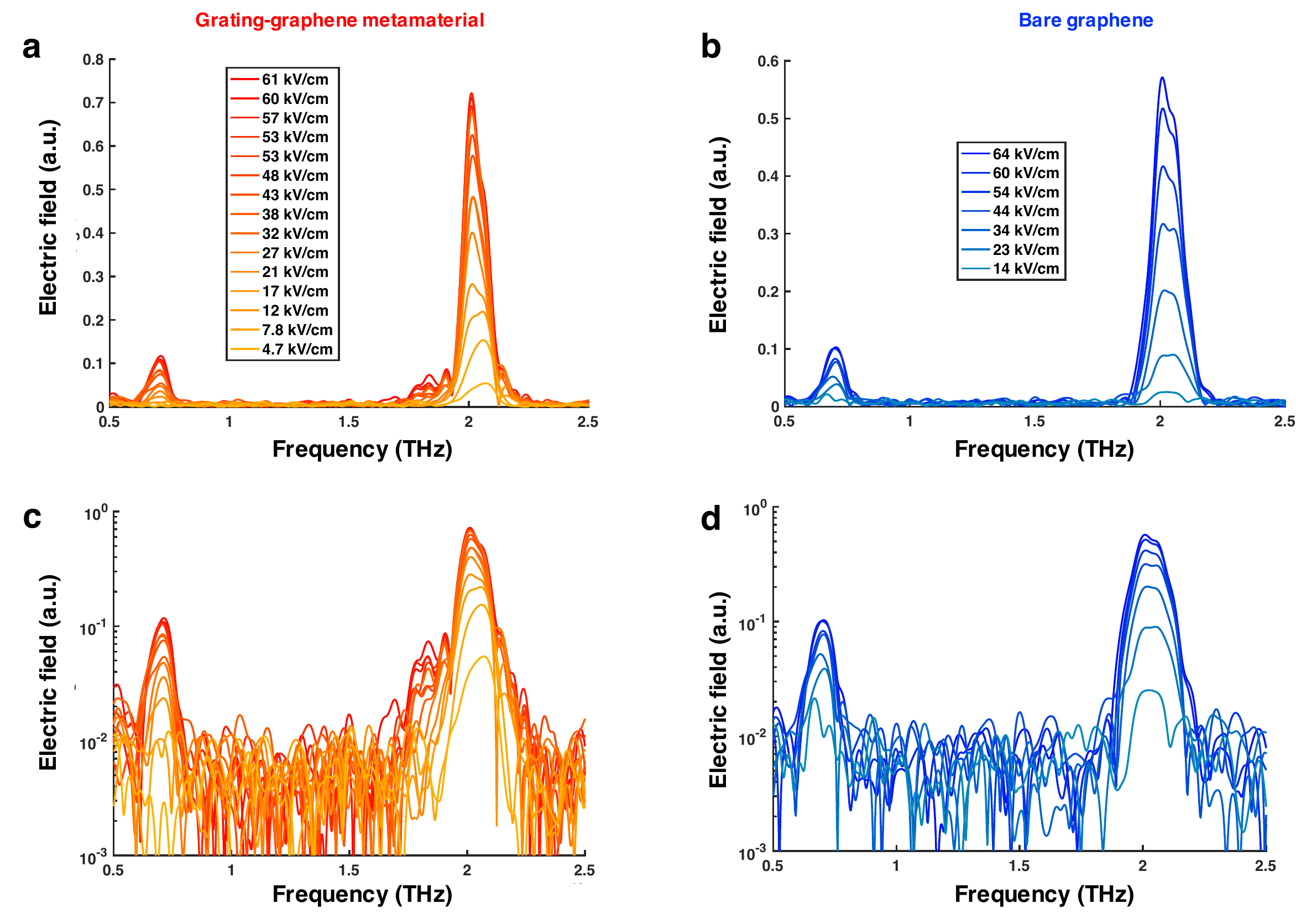}
   \caption{ \textbf{Fundamental and harmonic signals in frequency domain.}
The Fourier transforms of the measured time-domain electric field signals for grating-graphene metamaterial Sample A in linear \textbf{(a)} and logarithmic \textbf{(c)} scale; and for bare graphene Sample B in linear \textbf{(b)} and logarithmic \textbf{(d)} scale. There is a clear red shoulder on the data obtained with Sample A. 
}
\label{SuppFig2}
  \end{figure}

\end{document}